\documentclass[crop]{CSLB}

\usepackage{amsmath}
\usepackage{amsthm}
\usepackage{cite}
\usepackage{hyperref}
\usepackage{graphics}
\usepackage{algorithmic}
\usepackage{subfig}
\usepackage{url}

\usepackage{boites}

\jname{International Journal of Astrobiology}%

\newcommand{\ceti}{node}
\newcommand{\cetis}{nodes}
\newcommand{\hs}{\textsc{hearsay}}

\newcommand{\ffn}[1]{}

\newcommand{\ttn}[1]{}

\begin{document}

\supertitle{Research Paper}

\title[Probability of causal contacts]{Monte Carlo estimation of the probability 
of causal contacts between communicating civilisations}

\author[Lares, Funes \& Gramajo]{Lares M.$^{1, 2}$, Funes J.~G.$^{1, 3}$ \& Gramajo L.$^{1, 2}$}

\corres{\name{Lares M} \email{marcelo.lares@unc.edu.ar}}

\address{
   \add{1}{CONICET, Argentina}

   \add{2}{Universidad Nacional de C\'ordoba, Observatorio
           Astron\'omico de C\'ordoba, Argentina}

   \add{3}{Universidad Cat\'olica de C\'ordoba, Argentina}
}

\keywords{SETI, Computer simulations, Statistics}

%\JELclassification{XX; XX}
%\MSCcodes{XX; XX}

\Abbreviations{SETI: Search for Extraterrestrial Intelligence,\\
CCN: causally connected node,\\
SFC: surface of first contact,\\
SLC: surface of last contact,\\
DE: Discrete Event,\\
GHZ: Galactic Habitable Zone}

\begin{abstract}
    In this work we address the problem of estimating the
    probabilities of causal contacts between civilisations in the
    Galaxy. We make no assumptions regarding the origin and
    evolution of intelligent life.  We simply assume a network of
    causally connected nodes.  These nodes refer somehow to
    intelligent agents with the capacity of receiving and emitting
    electromagnetic signals. Here we present a three-parametric
    statistical Monte Carlo model of the network in a simplified
    sketch of the Galaxy. Our goal, using Monte Carlo simulations, is
    to explore the parameter space and analyse the probabilities of
    causal contacts. We find that the odds to make a contact over
    decades of monitoring are low for most models, except for those of
    a galaxy densely populated with long-standing civilisations. We
    also find that the probability of causal contacts increases with
    the lifetime of civilisations more significantly than with the
    number of active civilisations. We show that the maximum
    probability of making a contact occurs when a civilisation
    discovers the required communication technology.
\end{abstract}

\maketitle

\section{Introduction}\label{S_motivations}

The Drake equation  \citep{drake_intelligent_1962} provides a truly
helpful educated guess, a rational set of lenses --the factors in the
equation-- through which to look at future contacts with
technologically advanced civilisations in the Milky Way. 
The equation quantifies the number of civilisations from whom we might
receive an electromagnetic signal, using a collection of factors that
have been extensively discussed in the literature and whose estimated
values are revised continually. 
A comprehensive review and an analysis of each term of the equation are
presented in \citet{vakoch_drake_2015}.
Optimistic estimates from the Drake equation contrast with the
so-called Fermi paradox, which states the apparent contradiction between the
expected abundance of life in the Galaxy and the lack of evidence for
it \citep[e.g. ][]{hart_explanation_1975, brin_great_1983,
barlow_galactic_2012, forgan_galactic_2016, anchordoqui_is_2017,
Sotos_biotechnology_2019, carroll_nellemback_fermi_2019}.
There are many propositions aimed at solving this paradox, which make
use of statistical \citep{solomonides_probabilistic_2016, horvat_calculating_2007,
maccone_statistical_2015} or stochastic approaches
\citep{forgan_numerical_2009, bloetscher_using_2019,
glade_stochastic_2011, forgan_numerical_2010}.
Regarding the Drake equation, analytical interpretations
\citep{prantzos_joint_2013, smith_broadcasting_2009} or reformulations
\citep[][and references therein]{burchell_whither_2006} have also been
proposed. 
The absence of detections of extraterrestrial intelligent signals
could be explained by
astrophysical phenomena that makes life difficult to develop
\citep{annis_astrophysical_1999}.
Besides the possible scarcity of life,
alternative scenarios have also been discussed
\citep{barlow_galactic_2013, lampton_information_2013,
conway_three_2018, forgan_galactic_2017}.
The large distances in the Galaxy and the likely limited lifetime of
civilisations may play an important role in determining how difficult it
would be to obtain evidence for other inhabited worlds.
The analysis of these scenarios is difficult due to the lack of data
about the hypothetical extraterrestrial intelligences.
Indeed, as \citet{tarter_search_2001} pointed out, according to our
current technical capabilities for the search of extraterrestrial
intelligence (SETI), we have not received any signal yet.
The absence of detections has also motivated alternative ideas for new
SETI strategies \citep{forgan_exoplanet_2017, balbi_impact_2018,
loeb_eavesdropping_2006, maccone_KLT_2010, tarter_advancing_2009,
enriquez_breakthrough_2017, loeb_relative_2016, maccone_SETI_2011,
lingam_relative_2019, wright_theGsearch_2015, maccone_SETI_2013,
maccone_lognormals_2014, harp_application_2018,
forgan_possibility_2013, forgan_galactic_2017, funes_searching_2019}.

\Fpagebreak

The discussion about the problem of the unknown abundance of
civilisations in the Galaxy has been organised around the factors in
the Drake equation \citep{hinkel_interdisciplinary_2019}.
However, the uncertainties in these factors, specially the ones
representing biological processes, make it less suitable to a formal
study with the purpose of defining searching strategies or computing
the estimated number of extraterrestrial intelligences.
A number of studies propose alternative formalisms for the estimation
of the likelihood of detecting intelligent signals from space.
\citet{prantzos_joint_2013}, for example, proposes a unified framework
for a joint analysis of the Drake equation and the Fermi paradox,
concluding that for sufficiently long-lived civilisations,
colonisation is the most promising strategy to find other life forms.
\citet{haqq-misra_drake_2017} discuss the dependence of the Drake
equation parameters on the spectral type of the host stars and the
time since the Galaxy formed, and examine trajectories for the
emergence of communicative civilisations.
Some modifications to the original idea of the Drake equation have
been proposed, in order to integrate a temporal structure, to
reformulate it as a stochastic process or to propose alternative
probabilistic expressions.
Temporal aspects of the distribution of communicating civilisations
and their contacts have been explored by several authors
\citep{fogg_temporal_1987, forgan_spatiotemporal_2011,
balbi_impact_2018, balb_spatiotemporal_2018, horvat_impact_2011}, as
well as efforts on considering the stochastic nature of the Drake
equation \citep{glade_stochastic_2011}.

Several authors raise the distinction between causal contacts and
actual contacts.
In the first case, the determinations of metrics for the likelihood of
a contact are independent of considerations about the technological
resources or the implicit coordination to decipher intelligent
messages.
In a recent work, \citet{balbi_impact_2018} uses a statistical model to
analyse the occurrence of causal contacts between civilisations in the
Galaxy.
The author highlights the effect of evolutionary processes when
attempting to estimate the number of communicating civilisations that
might be in causal contact with an observer on the Earth.
\citet{cirkovic_temporal_2004} also emphasises the lack of temporal
structure in the Drake equation and, in particular, the limitations
of this expression to estimate the required timescale of a SETI
program to succeed in the detection of intelligent signals.
\citet{balbi_impact_2018} also investigates the chance of
communicating civilisations making causal contact within a volume
surrounding the location of the Earth.
The author argues that the causal contact requirement involves mainly
the distance between civilisations, their lifespan and their times of
appearance.
This is important since the time the light takes to travel across the
Galaxy might be much lesser than the lifetime of the emitter.
\citet{balbi_impact_2018} fixes the total number of civilisations and
explores the parametric space that comprises three variables, namely,
the distance to the Earth, the time of appearance and the lifespan of
the communicating civilisations.
Each of these three variables are drawn from a random distribution.
The distances are drawn from a uniform model 
for the
positions of civilisations within the plane of the Galaxy.
For the distribution of the characteristic time of appearance, the
author explores exponential and truncated Gaussian functions,
while for the lifespans chooses an exponential behaviour.
It is important to point out that the estimation of the number of
communicating civilisations vary with the choice of the statistical
model for the time of appearance, as shown by
\citet{balbi_impact_2018}.
For all analysed distributions, the author concludes that the
fraction of emitters that are listened is low if they are spreaded in
time and with limited lifetimes.
An analytical explanation of these concepts are presented 
in \citet{grimaldi_signal_2017}, who considers a statistical model for the
probability of the Earth contacting other intelligent civilisations,
taking into account the finite lifetime of signal emitters, and based
on the fractional volume occupied by all signals reaching our planet.

The quest for a formal statistical theory has also lead 
to important progress in the mathematical foundations of SETI.
Recently, \citet{bloetscher_using_2019} considers a Bayesian approach,
still motivated by the Drake equation, to estimate
the number of civilisations in the Galaxy.
To that end, the authors employ Monte Carlo Markov Chains over each factor
of the Drake equation, and combine the mean values to reach a probabilistic
result.
It is worth mentioning that the author proposes a log-normal target
distribution to compute the posterior probabilities.
This study concludes that there is a small probability that the Galaxy
is populated with a large number of communicating civilisations.
\citet{smith_broadcasting_2009} uses an analytical model to gauge the
probabilities of contact between two randomly located civilisations
and the waiting time for the first contact.
The author stresses that 
the maximum broadcasting distance and the lifetime of civilisations
come into play to produce the possible network of connections.

On this topic a number of works consider numerical simulations
\citep{forgan_evaluating_2015, vukotic_grandeur_2016,
murante_simulating_2015, forgan_numerical_2009, forgan_galactic_2017,
ramirez_new_2017}.
Although this approach does not rely on values that can be measured
from observations (like the fraction of stars with planets), it
depends on the definition of unknown or uncertain parameters required
to carry out the simulations.
In another numerical approach, \citet{vukotic_astrobiological_2012}
propose a probabilistic cellular automata modelling. 
In this framework, a complex system is modelled by a lattice of cells
which evolve at discrete time steps, according to transition rules
that take into account the neighbour cells.
The authors implement this model to a network of cells which represent
life complexity on a two--dimensional region resembling the Galactic
Habitable Zone (GHZ), an annular ring set between a minimal radius of
6~kpc and a peak radius of 10~kpc.
These simulations represent the spread of intelligence as an
implementation of panspermia theories.
Within this framework, \citet{vukotic_astrobiological_2012} also make
Monte Carlo simulations and analyse ensemble-averaged results.
Their work aims at analysing the evolution of life,
and although it does not account for the network of causal contacts
among technological civilisations, it
offers a tool to think SETI
from a novel point of view.

% INTRODUCTION OF *THIS* WORK

\begin{figure}[h]
\centering
\includegraphics[width=\columnwidth]{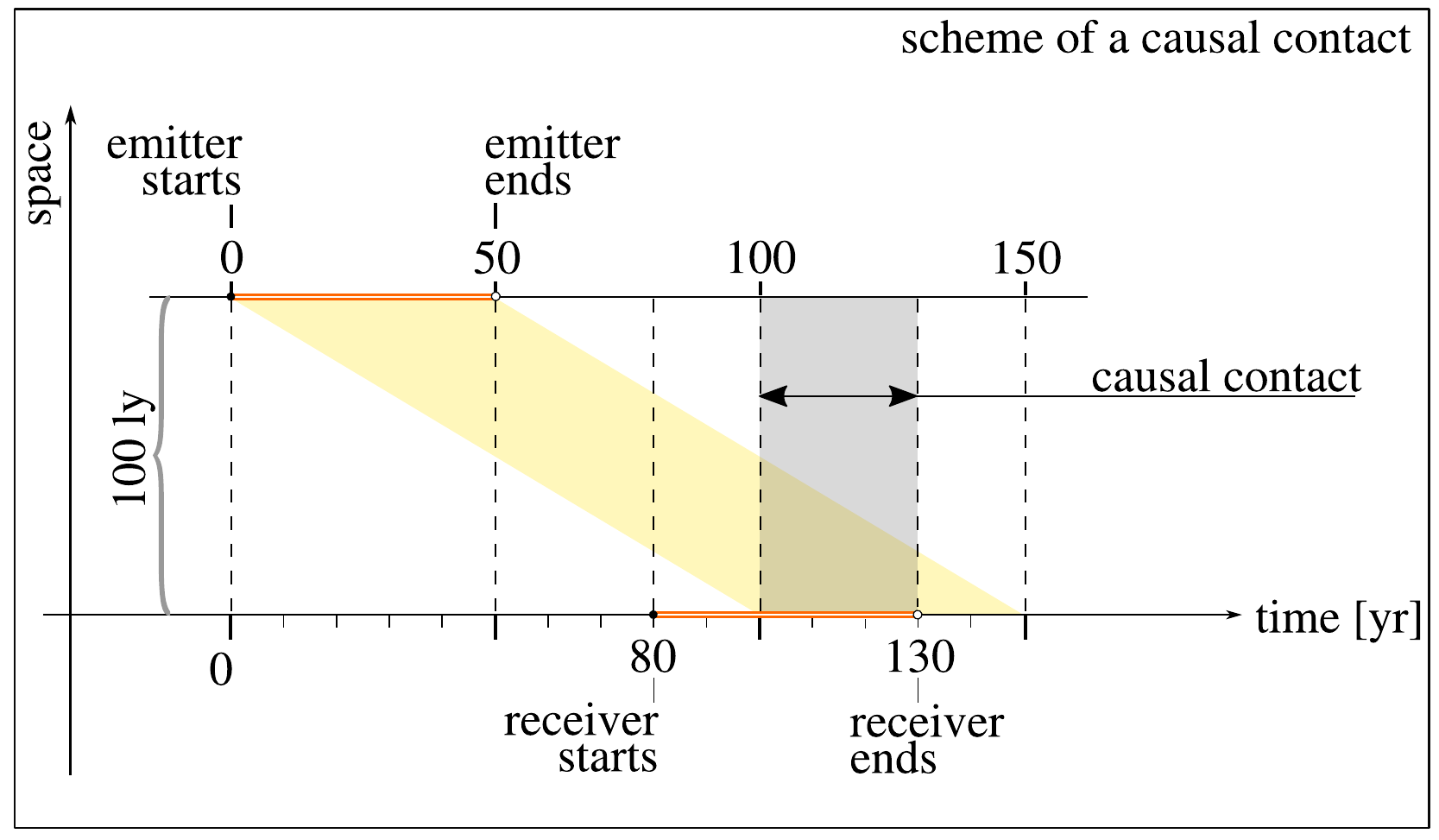}
\caption{Example of a configuration of an emitter and a receiver 
   to produce a causal contact.
   The observers are separated by 100~light years and appear with a
   difference in time of 80~years. One observer is active between
   $t=$0~yr and $t=$50~yr.  The other observer is also active for
   50~yr, between $t=$80~yr and $t=$130~yr. 
   The causal connection occurs between $t=$100~yr and $t=$130~yr.
   Meaningful times are marked with vertical dashed lines.
   }
\label{F_example}
\end{figure}

This study is conceived as an introduction to a simple probabilistic
model and a numerical exploration of its parameter space.
With these tools, we address the problem of the temporal and spatial
structure of the distribution of communicating civilisations.
This approach does not require asumptions about, e.g., the origin of
life, the development of intelligence or the formation of habitable
planets according to stellar type.
Instead, we assume monoparametric function families to model the
appearance of points in the disk of the Galaxy and over the time,
which we call \textit{nodes}.
These nodes represent the locations of ideal intelligent agents that
are able to receive and emit signals with perfect efficiency in all
directions.
Then we analyse the network of causally connected nodes limited to
have a maximum separation representing the maximum distance a signal
can travel with an intensity above a fixed threshold.
A node is causally connected to other nodes if it is within their light
cone.
We adopt as a definition that a light cone is the region in spacetime
within the surface generated by the light emanating from a given point
in space in a given period of time.
For example, two nodes separated by 100 light years which appear with
a difference in time of 80 years and last 50 years each, will be
causally connected for 30 years.
This connection is not, though, bidirectional.
In fact, the second node will see the first one for 30 years, but the
first node will never acknowledge the existence of the second.
An scheme of this example is shown if Fig.~\ref{F_example}.
This simple tought experiment exemplify the importance of the time
variable to analyse the structure of connections.
The model can be described with three parameters, namely, the mean
separation between the appearance of nodes, the mean lifetime of the
nodes, and the maximum separation to allow an effective causal
contact.
The model presented here comprises the stochastic networks of
constrained causally connected nodes, together with the three
parameters and additional hypotheses.
Hereafter, we refer to this model as the SC3Net model, standing for
''Stochastic Constrained Causally Connected Network``.

We provide a framework to explore, through a suite of
numerical simulations, the parameter space of three unknown
observables.
This allows to discuss possible scenarios and their consequences in
terms of the probability of making contacts.
The method we use for simulating a stochastic process is an
approximation that allows to study the behaviour of complex systems,
by considering a sequence of well defined discrete events.
The simulation is carried out by following all the variables that
describe and constitute the state of the system.
The evolution of the process is then described as the set of changes
in those variables. 
In this context, an event produces a specific change in the state that
can be triggered by random variables that encode the stochastic nature
of the physical phenomenon.
For example, when a new contact is produced between two entities in
the simulated galaxy, the number of active communication lines is
increased by one.
Also, if it is the first contact for that nodes, then the number of
communicated nodes increases by two.
When a new node becomes active, the sistem has an increase of one in
the number of nodes, although the number of communcations does not
necessarily change.
The process then involves following the changes on the state of the
system, defining the initial and final states.
This is done by defining a method that allows to keep track of the
time progress in steps and maintaining a list of relevant events,
i.e., the events that produce a change in the vriables of interest.
With this method we abandon the frequentist approach of the Drake
equation to compute the number of civilisations, providing instead its
statistical distribution.
More importantly, we are interested on the probability of contacts,
which depend critically on the time variable.
This is an exploratory analysis that aims at developing a numerical
tool to discuss the different scenarios based on statistical
heuristics.
The approach proposed here should be considered as a compromise
between the uncertainties of the frequentist estimations and the
detailed recipes required on the numerical simulations.
It is worth noticing that the SC3Net model is not intended for a
formal fit at this stage due to the lack of data, but it can help to
understand how unusual it would be to actively search for intelligent
signals for 50 years without possitive detections.

This paper is organized as follows.
In Sec.~\ref{S_methods} we introduce the methods and discuss the
candidate distributions for the statistical aspects of the times
involved in the communication process.
We present our results in Sec~\ref{S_results}, with special emphasis
on the statistical distributions of the duration of causal contacts in
and the distribution of time intervals of waiting for the first contact.
This quantities are considered as a function of the three simulation
parameters.
In Sec.~\ref{S_discussion} we discuss our results and future research
directions.

\begin{figure}[!t]
   \centering
   \includegraphics[width=0.5\textwidth]{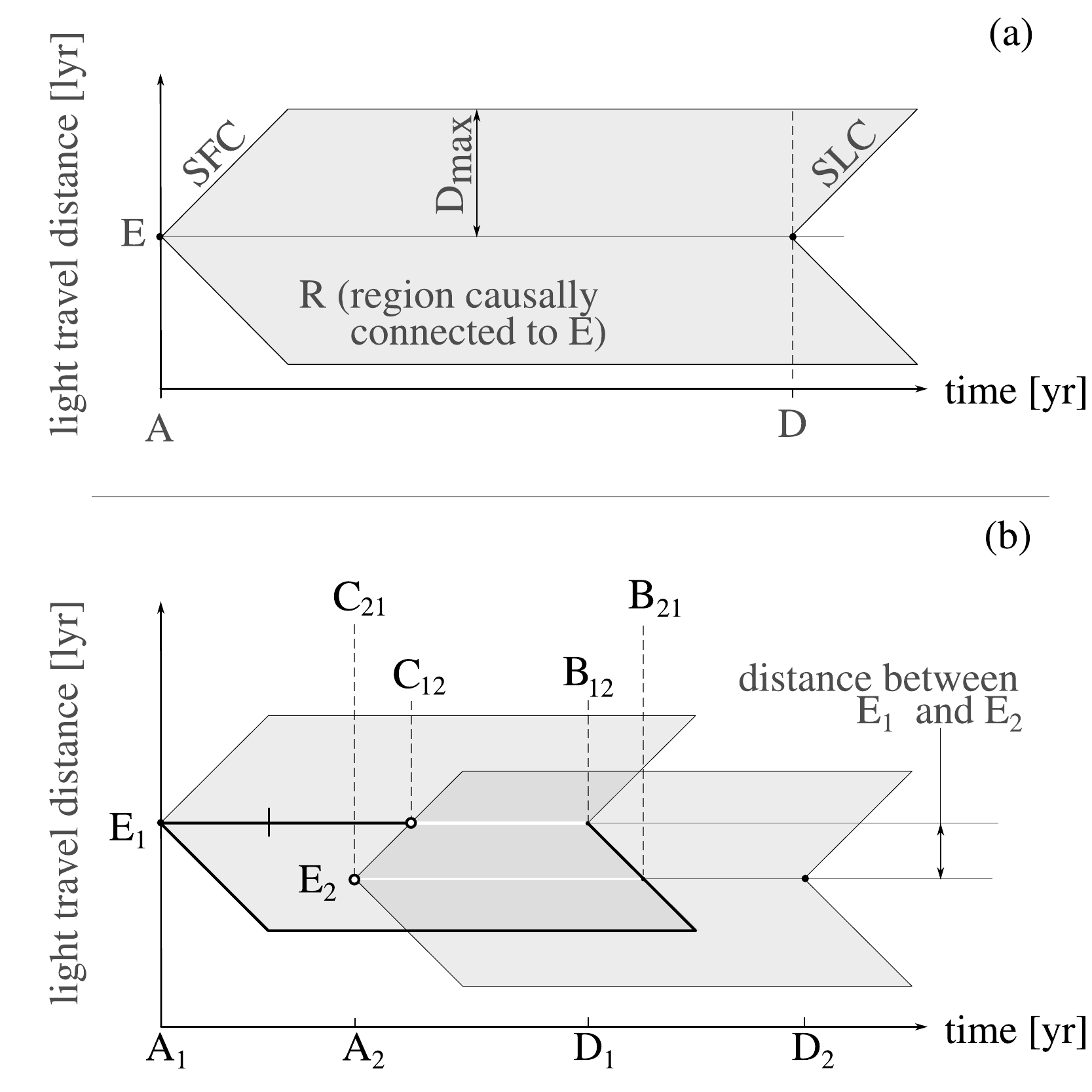}
   \caption{
Space--time diagrams showing a schematic representation of the different stages
in the development of a constrained causal contact node.
Panel (a) represents the region $R$ in space and time which is causally
connected to the emitter (E, left vertex), following an ``A'' type
event (``Awakening'') in which the node acquires the communication
capability.
The sphere of the first contact (SFC) is the sphere centred on the emitter that
grows until its radius reaches the D$_{\rm max}$ distance, in which the
power of the signal would equal the detectability threshold.
This sphere is represented by the left triangle of the region
$R$.
The surface of the last contact (SLC) is another sphere that grows from a ``D''
event (``Doomsday'').
The region which is causally connected to the emitter is then limited by these
two spheres, and has the shape of a sphere or of a spherical shell,
depending on the time.
The temporal intervals for the communication between two \cetis{} are
represented in the panel (b).
The node E$_1$ can listen to signals from the node E$_2$, since
the ``Contact'' event (t=C$_{12}$) up to the ``Blackout'' event
(t=B$_{12}$).
Similarly, the node E$_2$ is in causal contact from the
   node E$_1$, in the time interval (C$_{21}$, B$_{21}$).
} \label{F_scheme}
\end{figure}

\section{Methods and working hypotheses}\label{S_methods}

Simulations are suitable tools to analyse systems that evolve with time and
involve randomness.
An advantage of a numerical approach is that it usually requires fewer assumptions
and simplifications, and can be applied to systems where analytical models are
hard or impossible to develop.
In particular, a suitable tool to model complex stochastic processes through
the changes in the state of a system is the discrete--event (hereafter, DE)
simulation approach.
A system described with the DE paradigm is characterised by a set of actors and
events.
Actors interact causally through a series of events on a timeline and process
them in chronological order \citep{ptolemaeus_system_2014,
chung_simulation_2003, ross_simulation_2012}.
Each event modify the variables that define the process, producing the
corresponding change in the state of the simulated system.
This method is well suited for the particular case of the diffusion of
intelligent signals in the galaxy and allows to explore several models easily.
We simulate the statistical properties of a set of points in space and time
that have a causal connection at light speed and are separated
by a maximum distance.
We refer to this nodes as ''Constrained Causal Contact Nodes``
(hereafter, \ceti{}).
We choose this generic name in order to stress the fact that in this analysis 
only the causal contact is considered, independently of any broadcasting or
lookout activity.
The system we propose is ideal, in the sense that it considers the special case
of a fully efficient node that emits and receives isotropically.
A causal contact node, considered as a broadcasting station that has the
ability to detect signals through an active search program, could have a
lesser than one efficiency factor, which is not included in the current
analysis.
Also, it is worth mentioning that this is a general approach, and not
necessarily a \ceti{} is the host of an intelligent civilisation.
It can be associated with a planet where life has developed,
became intelligent, reached the skills required to find the right communication
channel, sustained a search and established a contact.
Alternative message processing entities could be considered, for example
interstellar beacons where intelligent life has ceased to exist but continue with its
emission, or communication stations established by probes or left by
intelligent beings \citep[see, e.g., ][]{peters_outer_2018,
barlow_galactic_2013}.
In principle, these strategies could affect our results since it would be
easier to configure a cluster of \cetis{} that spread in time.
However, we do not consider these speculative alternatives at this point.
For the purpose of the present analysis, only the communication capability is
relevant, since we study the causal contacts between the locations.
The system is defined by a number of actors that represent \cetis{} and appear
at different instants in time, generating events that produce meaningful
changes in the variables that describe the system, i.e., in the arrangement of
\cetis{} and their network of causal contacts.
For example, the appearance of a new \ceti{} in a region filled with a signal
emitted by another node, will increase the number of active \cetis{} and the
number of pairs of \cetis{} in causal contact.
Assuming some simple hypotheses, the discrete events method can be performed
taking into account a small number of variables, which allow to analyze the
variation of the results in the SC3Net model parameter space.

In what follows, we outline the experimental setup adopted to estimate the
probabilities of causal contacts and several derived quantities.
This is done in terms of three independent parameters, namely, the mean time
span between the appearance of consecutive \cetis{}, $\tau_{\rm a}$,
their mean lifetime $\tau_{\rm s}$, and the maximum distance a signal
can be detected by another \ceti{} ($D_{\rm max}$).
Intuitively, the probability of the existence of causal contacts between pairs
of \cetis{} would be larger for smaller $\tau_{\rm a}$ parameters,
higher $\tau_{\rm s}$ or
higher $D_{\rm max}$ parameter values.
We also propose theoretical distribution functions for both the lifespans
($\tau_{\rm s}$) and the number of \cetis{} per unit time
\citep{maccone_evolution_2014, Sotos_biotechnology_2019}.
The later is related to the time span between the appearance of consecutive
\cetis{} (since when $\tau_{\rm a}$ is shorter, it produces a greater
density of \cetis{}).
The analytical expresions for these distributions are set to a fixed law, as
discussed in Sec.~\ref{SS_PDF_shape}.

\ffn{1} Space--time diagrams, where time and space are represented on
the horizontal and vertical axes, respectively, are suitable tools to
describe the causal connections among different nodes.
We illustrate in the Fig.~\ref{F_scheme} the schematic representation
of the region causally connected to a given \ceti{}.
For the sake of simplicity, we show in the plot the light travel
distance, i.e., the distance traversed by any signal spreading from
the emitter at light speed (for example, any electromagnetic signal).
In this scheme, a light pulse would follow a trajectory represented by
a line at 45 degrees from the axes, given the units of time and
space axes are years and light years, respectively.
Panel (a) represents the region $R$ (shaded polygon)
which is causally connected to the emitter (left vertex).
This region develops after an event (hereafter dubbed A-type event) in
which the node acquires the communication capability, or becomes
''active''.
The sphere of first contact (SFC) is centred on the emitter,
represented by the left angle of the region $R$.
This sphere grows until its radius reaches the D$_{\rm max}$ distance, in
which it would be no longer detectable due to the decrease in the
energy per unit area, which falls under the assumed detectability
threshold.
Similarly, the surface of last contact (SLC) is another sphere that
grows from an event in which the nodes ceases to possess the
communication capability (D type event), and carries the last signal
produced by the emitter.
The region which is causally connected to the emitter is then limited
by these two spheres and therefore has the shape of a filled sphere
(before the D event) or of a spherical shell (after the D event).
The temporal intervals for the communication between two \cetis{} are
represented in the panel (b) of the Fig.~\ref{F_scheme}.
In this scheme, the entire lifetime of a \ceti{} can be represented by
a polygon.
This polygon is limited in the spatial axis by the maximum distance of
the signal $D_{\rm max}$ and in the time axis by the time span between the
A-type event (concave vertex) and the D-type event (convex vertex).
This representation allows to visualize the important events that
result from the presence and communications of two \cetis{}.
The points in time where the \cetis{} acquire their communicating
capacity are dubbed ``Awakening'' events ($A_1$ and $A_2$).
Similarly, the instants in time where the \cetis{} lose their
communicating capacity are dubbed ‘’Doomsday’’ events ($D_1$ and $D_2$).
The points in space--time where the first contact is produced for each
one of the \cetis{}, are defined as ``Contact'' events, shown as $C_1$
and $C_2$.
Finally, the two points where the contact is lost for each one the
\cetis{}, are denominated ``Blackout'' events, presented as $B_1$ and
$B_2$.
The receiver node E$_1$ can listen to signals from the emitter node
E$_2$, from the first contact event (t=C$_{12}$) until the last
one (t=B$_{12}$).
Similarly, the receiver node E$_2$ can listen to signals from the emitter
node E$_1$, from t=C$_{21}$ up to t=B$_{21}$.
The time intervals for the open communication channel are determined
by the ``C'' and ``B'' type events.
It is noteworthy the fact that there is a time delay between the
contact events of the two \cetis{} involved in this analysis, and also
a time delay between the two blackout events.
Therefore, the time range when a bidirectional contact is possible
occurs between the maximum time of the contact points and the minimum
time between the blackout points.

The temporal structure that emerges from the experimental setup implies that,
as a fundamental property of the simulations, a causal connection can be
produced without  requiring that the two nodes are active at the same time.
This property of the system arises as a consequence of the large spatial and
temporal scales, where a message is transmitted at the (relatively small) light
speed.
Although a \ceti{} could be active for a large enough period to transmit a
message at large distances, the limited power of the message and the dilution
that depends on the squared distance from the source imposes a detectability
limit.
As a consequence of this limitation and of their finite lifetime, considered as
the period between the acquisition and loss of communicating capacity, each
\ceti{} will fill a spherical shell region of the galaxy, limited by two
concentric spherical surfaces.
The leading front, or surface of first contact (SFC) grows from the central
\ceti{} until it reaches the maximum distance $D_{\rm max}$.
Following the end of the civilisation, there is still a region which is filled
with the emitted signals.
This approach has been also considered in other statistical models \citep[e.g.,
][]{smith_broadcasting_2009, grimaldi_signal_2017, Grimaldi2018}.
The trailing front, or surface of last contact (SLC) also grows from the
central \ceti{}, with a delay with respect to the SFC equivalent to the
lifetime of the node, and produces a spherical shell region.
Any other \ceti{} within this region will be in causal contact with the originating node,
even if it has disappeared before the time of contact.
This region will grow if the surface of first contact has not yet reached the
maximum distance $D_{\rm max}$, and will shrink after a D-type event until the
surface of last contact reaches $D_{\rm max}$, producing as a result the loss
of all signal from the central \ceti{}.
In our approach, we consider a model galaxy where the width of the disk is
negligible with respect to the radius of the disk.
In the 2D simulation only the intersection of the communicating spherical
shells with the plane of the galaxy is relevant, and produce the corresponding
circles or rings for the filled spheres or annular regions of the spherical
shells, respectively.
The initially growing communicating sphere is shown over space--time diagrams,
where time is represented on the horizontal direction, and space is represented
in the vertical direction.
In the Fig.~\ref{F_scheme} the two emitters in panel (b), E$_1$ and E$_2$,
reach each other at different times.
The time span for E$_i$ is (A$_i$ , D$_i$) , for $i = {1, 2}$.
Emitter $i$ can listen to emitter $j$ between C$_{ij}$ and B$_{ij}$.
The type and length of causal contact in both directions depend on the distance
and time lag between the awakening events, the maximum distance that a signal
can reach and the time period in which each emitter is active.

In our experimental configuration the simulation starts assuming that the
stochastic process is already stable, and finishes before any galactic
evolution effect could modify the fixed values of the variables.
Likewise, we assume that the probability for the appearance of a \ceti{} is
homogeneous over the GHZ.
The adopted geometry of the GHZ in all the simulations is given by a 
two-dimensional annular region,
with an inner radius of 7~kpc and an outer radius of 9~kpc
\citep{lineweaver_galactic_2004}.
Although the Galaxy has a well-known spiral structure, the nodes are assumed
to be sparse (otherwise the Galaxy would be full of life) and the spiral
structure would not, in that case, produce significant differences.
If the distribution of nodes is not sparse, as it could be the case if the
spiral arms host most of the \cetis{}, then contacts would be more frequent
between closely located nodes.
In such a case our results underestimate the number of contacts between
close pairs of nodes within the same spiral arm, and conversely, overestimate
the number of contacts between separated nodes.
We also limit the possibilities of life or other types of civilisations to the
usually stated hypotheses for the definition of the GHZ
\citep{dayal_habitability_2016, gonzalez_galactic_2001,
lineweaver_galactic_2004, gonzalez_habitable_2005, morrison_extending_2015,
haqq-misra_evolution_2019, rahvar_cosmic_2016, gobat_evolution_2016,
rahvar_cosmic_2016} and consider that habitability remains constant over time
\citep[see, however, ][]{gonzalez_habitable_2005, dayal_habitability_2016,
gobat_evolution_2016}.
This means that we set aside civilisations that could survive in severe
conditions or unstable systems, which would prevent the appearance of life as
we know it.

We stress the fact that we are
considering causal contacts instead of actual contacts.
In more realistic scenarios, there are several sources of ``signal loss'' with
respect to the ideal case.
Among them, we can mention temporal and signal power aspects and direction
dependent communication capabilities.
The results must then be interpreted as the case of ideal nodes, with a perfect
efficiency in the emision and reception of signals.
The probabilities of contacts presented here are then upper limits to the
number of communications between civilisations in the Galaxy, given the
implemented hipotheses in the model.
The use of light cones as causal contact regions is inspired by the fact that
light-speed traveling messengers like electromagnetic radiation, gravitational
radiation or neutrinos are often considered as possible message carriers
\citep{hippke_interstellar_2017, wright_how_2018}.
This excludes messages sent with mechanical means or physical objects
\citep[e.g., ][]{Armstrong2013, barlow_galactic_2013}, or through some unknown
technology that violates the known laws of physics.

As part of this benchmark, we assume that the capacity to emit and receive
signals occur at the same time.
Although there are several reasons to think that this could not be the case, at
large time scales it can be considered that both abilities occur roughly at
coincident epochs.
In the ideal setup, this would be equivalent to nodes that send messages
isotropically and scan the local skies on all directions with a perfect
efficiency.
Another essential assumption is that all \cetis{} use the same signal power, so
that there is a maximum distance out to which it can be detected.
In such a system we compute probabilities of a random node making contact with
another node, i.e., they are not specific for the case of contacts with the
Earth.
Regarding the extent of the signals, we know that the distance from which a
signal from Earth could be detected using the current technology is about few
parsecs, given that the signal was sent to a specific direction.
It is straightforward to propose and implement a distribution of maximum
distances, although this would increase the model complexity at the cost of a
larger uncertainty.

In our simulations, we assume the simplest configuration for the
growth of the sphere of first contact.
In particular, we do not consider the
possibility of stellar colonisation
\citep[e.g.][]{newman_galactic_1981, walters_interstellar_1980,
starling_virulence_2013, barlow_galactic_2012, jeong_large_2000,
maccone_mathematical_2011}.
We also assume that communication is
equally likely in all directions, i.e., we assume isotropic
communication in all cases.
Different communication efficiencies or
detection methods are straightforward to carry out in the simulations
for more detailed and complex scenarios.
This approach, however, is
beyond this work because it obscures the experimental setup and make
the results less clear.
The assumptions we accept imply that the
results are independent of whether intelligent agents are organic or
artificial.
Moreover, the causes of the limited lifetime of a
civilisation can be natural (astrophysical phenomena), caused by auto
destruction or by external factors, to name a few.
However, we assume
that these events are sufficiently numerous in the Galaxy so that a
statistical model is plausible. The failure of this hypothesis would
imply extreme values for the parameters that represent the density of
the nodes (i.e., $\tau_{\rm a}$).

 \setlength{\tabcolsep}{10pt}
\begin{table*}
\centering
\begin{tabular}{cllll}
\hline
   \multicolumn{2}{l}{independent variables (parameter space)}
   &min. value&max. value&Nbins\\
\hline
   $\tau_{a}$ & Mean temporal separation between consecutive awakenings 
	      & 5$\cdot10^2$~yr & 40500~yr & 21\\ 
	    & (linear grids) & 5$\cdot$10$^5$~yr & 10$^6$~yr & 51\\ 
   $\tau_{s}$ & Mean lifetime of a \ceti{}
	      & 5$\cdot10^2$~yr & 40500~yr & 21\\ 
	    & (linear grids) & 5$\cdot$10$^5$~yr & 10$^6$~yr & 51\\ 
	D$_{\rm max}$ & Maximum reach of a message  & \multicolumn{3}{c}{100 pc, 500 pc, 1 kpc, 5 kpc, 10 kpc} \\
\hline
   \multicolumn{2}{l}{fixed variables} & \multicolumn{2}{c}{assumptions} &value \\
\hline
   & statistical properties of all \cetis{} &equally distributed&&\\
   & Point process for the distribution in time & homogeneous &&\\
   f$_s$ & The scan of the sky & fully efficient&&1\\
   f$_p$ & panspermia or colonization &absent&&0\\
   & shape of the Galactic Habitable Zone & two-dimensional ring &&\\
	R$_{GHZ}^{\rm min}$   & Inner radius of the GHZ  & \citet{lineweaver_galactic_2004} & & 7~kpc\\
   R$_{GHZ}^{\rm max}$   & Outer radius of the GHZ       & \citet{lineweaver_galactic_2004} & & 9~kpc\\
	t$_{\rm max}$ & Time span of the simulation  & &
   \multicolumn{2}{r}{10$^5$~yr - 10$^9$~yr} \\
    & \multicolumn{3}{l}{number of random realizations for each point
    in the parameter space} & 1 - 50 \\
\hline
   \multicolumn{2}{l}{discrete events} &
   \multicolumn{3}{l}{affected variables}\\
\hline
   A event & Awakening: a \ceti{} starts its communication
   capabilities &Number of active \cetis{}\\
   B event & Blackout: the end of the communication channel stops
   &Number causal channels\\
   C event & Contact: a new causal contact is produced &Number of causal channels\\
   D event & Doomsday: a \ceti{} ends its communication capabilities&Number of active \cetis{}\\
\hline

\hline
\end{tabular}
\caption{Definition of independent variables and adopted values for 
   fixed parameters 
   that are part of the simulation.  Variable parameters define the
   spatial and temporal structure
   of the process and the maximum reach of the messages.}
\label{T_simu_hypotheses}
\end{table*}

\subsection{Power laws vs. exponential laws}\label{SS_PDF_shape}

The temporal structure of the process is defined by two distribution
parameters. One of them represent the mean time interval that a node can
emit and receive signals (its lifetime), and the other the mean time interval between the
emergence of consecutive \cetis{}.
The spatial structure of the simulation is
given by the size and shape of the Galactic Habitable Zone and the maximum
distance a signal can travel to be detected (D$_{\rm max}$).
The parameters for
the temporal distributions also determine the spatial properties, since the
density of active \cetis{} in the galaxy depend on these two parameters.
For
example, small $\tau_{\rm a}$ and a large $\tau_{\rm s}$ will produce a densely
populated galaxy (in this context, a galaxy is an element in a 
statistical ensemble, not the Milky Way).
Also, some hypotheses regarding the shapes of the
distributions of the temporal parameters must be made in order to complete the
simulation.
\citet{forgan_spatiotemporal_2011} argues that the times at which
different civilisations become intelligent follow a Gaussian distribution, and
then the distribution of inter-arrival times is an inverse exponential.
We
assume that the distribution of the times of A events is a stationary Poisson
process, and then the distribution of the times between the appearance of new
consecutive nodes is exponential.
Regarding the duration of a node, we
propose that its distribution is a stationary
exponential distribution.
There are no clear arguments to conclude a
statistical law for the later distribution, so that we propose it as a working
hipothesis.  
This heuristic does not make any consideration about the origin of life, although
different approaches are possible.
For example,
\citet{maccone_lognormals_2014} argues that this distribution should be a log-normal. 
Preferentially, a theoretical statistical distribution of the lifespan of
civilisations would rely on the basis of the underlying
astrophysical and biological processes \citep{balbi_impact_2018}.

The power law and exponential statistical distributions are among the most
common patterns found in natural phenomena.
For example, the distribution of the frequency of words in many languages is
known to follow the law of Zipf (which is a power law).
These distributions arise any time a phenomenon is characterised by commonly
occurring small events and rarely produced large events \citep[e.g.
][]{adamic_zipf_2000}.
Zipf law also describes population ranks of cities in various countries,
corporation sizes, income rankings, ranks of the number of people watching the
same TV channel, etc.
The magnitudes of earthquakes, hurricanes, volcanic eruptions and floods; the
sizes of meteorites or the losses caused by business interruptions from
accidents, are also well described by power laws
\citep{sornette_critical_2006}.
Additional examples include stock market fluctuations, sizes of computer files
or word frequency in languages \citep{mitzenmacher_brief_2004,
newman_power_2005, simkin_theory_2006}. 
Power laws have also been widely used in biological sciences.
Some examples are the analysis of connectivity patterns in metabolic networks
\citep{jeong_large_2000} or the number of species observed per unit area in
ecology \citep{martin_origin_2006, frank_common_2009}.
More examples can be found in the literature \citep{martin_origin_2006,
maccone_KLT_2010, barabasi_scale_2009, maccone_evolution_2014, maccone_lognormals_2014}.
This distribution family is suitable for the statistical description
of the $D_{\rm max}$ parameter, although in this work we assume a uniform
distribution for simplicity.
The power law family of functions is also a good candidate for the description of
the temporal variables in the model.
However, we prefer the exponential model.
The exponential distribution of lifespan and waiting times is justified by
considering the hypothesis that the process of appearance of life in the galaxy
is homogeneous and stationary.
That is, there is not a preferred region within
the GHZ for the spontaneous appearance of life, and the emergence of a node is
independent of the existence of previous nodes in the galaxy.
These seem to be simple conditions, and allow to propose a distribution family
without knowing the details of the underlying process.
The exponential distribution for the separations in time is equivalent to
proposing a Poisson process for the emergence of nodes, given the relation
between the number of events in time or space and the waiting time or
separation, respectively \citep[e.g., ][]{ross_simulation_2012}.
That is, these
are two alternative approaches to describing the same process, a Poisson
distribution for the number of events implies an exponential distribution for
their separations, and vice versa.
It should be emphasised that the exponential
laws used in this work are assumed as part of the working hypothesis, and
instead of analysing results from a particular parameter chosen ad hoc, we
explore the parameter space and analyse the impact of the values of these
parameters on the results.

\begin{figure} \centering
   \includegraphics[width=0.5\textwidth]{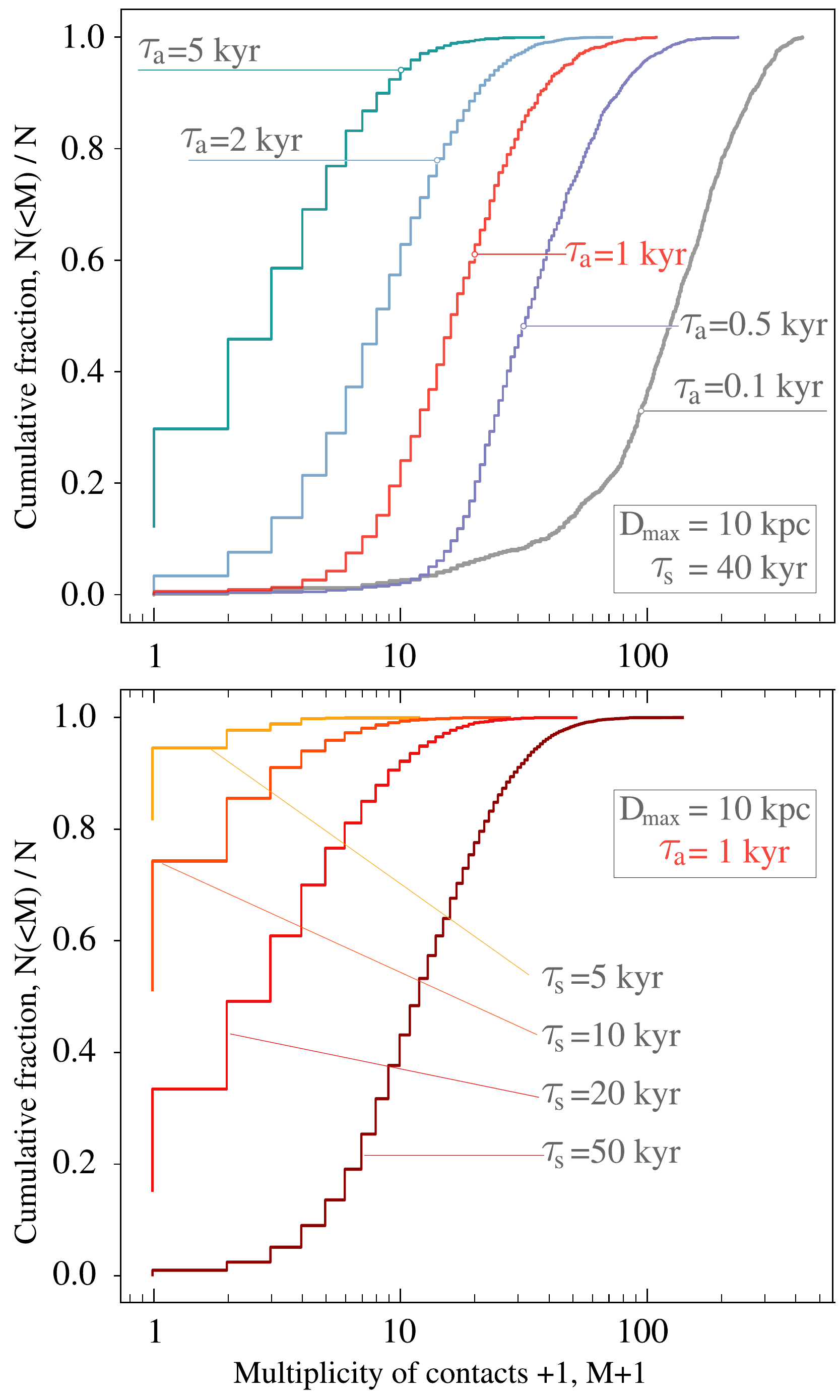}
   \caption{
Empirical cumulative distributions of the number contacts
   for \cetis{} in different samples and with D$_{\rm max}$=10~kpc.
The upper panel shows the variation of the distribution
   as a function of the $\tau_{\rm a}$ parameter, including the values 
   0.1, 0.5, 1, 2, and 5~kyr. All the curves
   correspond to models with 
   $\tau_{\rm s}=$40~kyr and \mbox{D$_{\rm max}=10$~kpc}.
   The bottom panel corresponds to 
   $\tau_{\rm a}=1$~kyr and D$_{\rm max}=10$~kpc, and the following values
   of $\tau_{\rm s}$: 5, 10, 20 and 50~kyr.
   Cumulative distributions can be used to visualise the probability
   of a random node of having more than M contacts in a given model.  For
   example, in the upper panel the probability of a random node having
   more than 10 contacts is nearly one for the model with the
   smallest $\tau_{\rm a}$ value, and nearly zero for the model with the
   largest $\tau_{\rm a}$ value.
   } \label{F_number_of_contacts}
\end{figure}

\subsection{Complexity of the model}

In this Section we discuss the degree of complexity in the model, considering a
compromise between the accuracy of the model and the number of parameters that
are free or with a high uncertainty.
Firstly, we emphasise that the odds of a causal contact between two nodes
should not be considered as the odds of a contact between two intelligent
civilisations, and in fact the latter could be much lesser than the former.
Indeed, in order to establish a contact between any two entities, a minimum
degree of compatibility must be accomplished without any previous agreement,
making the possibility of a contact with a message that could be deciphered
highly rare \citep[see e.g. ][]{forgan_collimated_2014}.
Besides the trade--off between the simplicity and the complexity of the
experiments, further analysis could be performed following this framework in
order to explore possible implications of the results for more detailed
configurations.
For example, the communication method (isotropic, collimated, serendipitous)
can affect the observables, making it necessary to implement a correction
factor.
Taking this into account, our results regarding the probability of causal
contact should be considered as upper limits for effective contacts, since they
depend on the efficiency of both the emitter and the receiver to broadcast and
scan the sky for intelligent signals, respectively
\citep{grimaldi_signal_2017}.
A correction by a coverage ratio in the detection and by a targeting ratio in the
emission could be easily implemented in the simulation, although the effect of
reducing the probability of contact is basically the product of the efficiency
ratios and thus such implementation is not necessary.
Therefore, the values of the probabilities could be modified by a constant
correction factor equal to the combined emission/reception efficiency
\citep{smith_broadcasting_2009, anchordoqui_upper_2019, forgan_collimated_2014}
or for beam-like transmissions \citep{grimaldi_signal_2017}. 
Other considerations include the effects of alignments, the use of stars as
sources or amplifiers \citep{Edmondson2003, borra_searching_2012}, or the
nature of the message carrier.
Another improvement could be the use of a spatial distribution that resembles
the spiral shape and the width of the disc of the Galaxy.
Regarding the distributions of the parameters, different distribution functions
for the mean lifetime of \cetis{} can be implemented as alternatives.
Regarding the degree of idealisation, the model admits different
efficiencies for the sphere of the causal region of each \ceti{}, featuring
different searching strategies \citep{hippke_interstellar_2017}. 
It is also possible to consider that $D_{\rm max}$ is different for different
\cetis{}.
For example, a power law where a powerful emission is rare and a low emission
is common could be an improvement to the model.
In this work we choose not to implement this for the sake of simplicity.
Finally, the role of the message contents could influence on the lifespan of a
\ceti{} that receives a message, although the implementation of such behaviour
would increase the number of free parameters and would be more speculative.   
We limit the scope of this work to a simple version of the model. 
Once the model has been defined, it can be implemented as a discrete event
simulation, as described in detail in the next Section.

\subsection{Discrete event process}

A discrete event simulation is performed for a given model, in this
work $M(\tau_{\rm a}, \tau_{\rm s},
D_{\rm max})$, by keeping track of a set of variables that change each time an
event happens. 
In addition, the model comprises elements that are fixed for all simulations,
for example, the functional forms of the statistical distributions and the
adopted values of particular variables (see \ref{T_simu_hypotheses}).
The main variables that follow the evolution of the simulation are:
the positions of stars, which are sampled randomly within the GHZ; the time of
the awakening of each node (A event); and its time of disappearance (D event).
The variables that can be deduced from the previous ones include: the number of
\cetis{} in casual contact with at least another \ceti{} at a given time; the
number of \cetis{} as a function of time; the number of \cetis{} that receive at
least one message; the number of \cetis{} that receive a message at
least one time and successfully deliver an answer; and the number distribution
of waiting times to receive a message.
All these quantities are updated each time one of the four events (A, B, C, D)
occurs.   

\begin{figure*} % 2D color plot
   \centering
   \includegraphics[width=\textwidth]{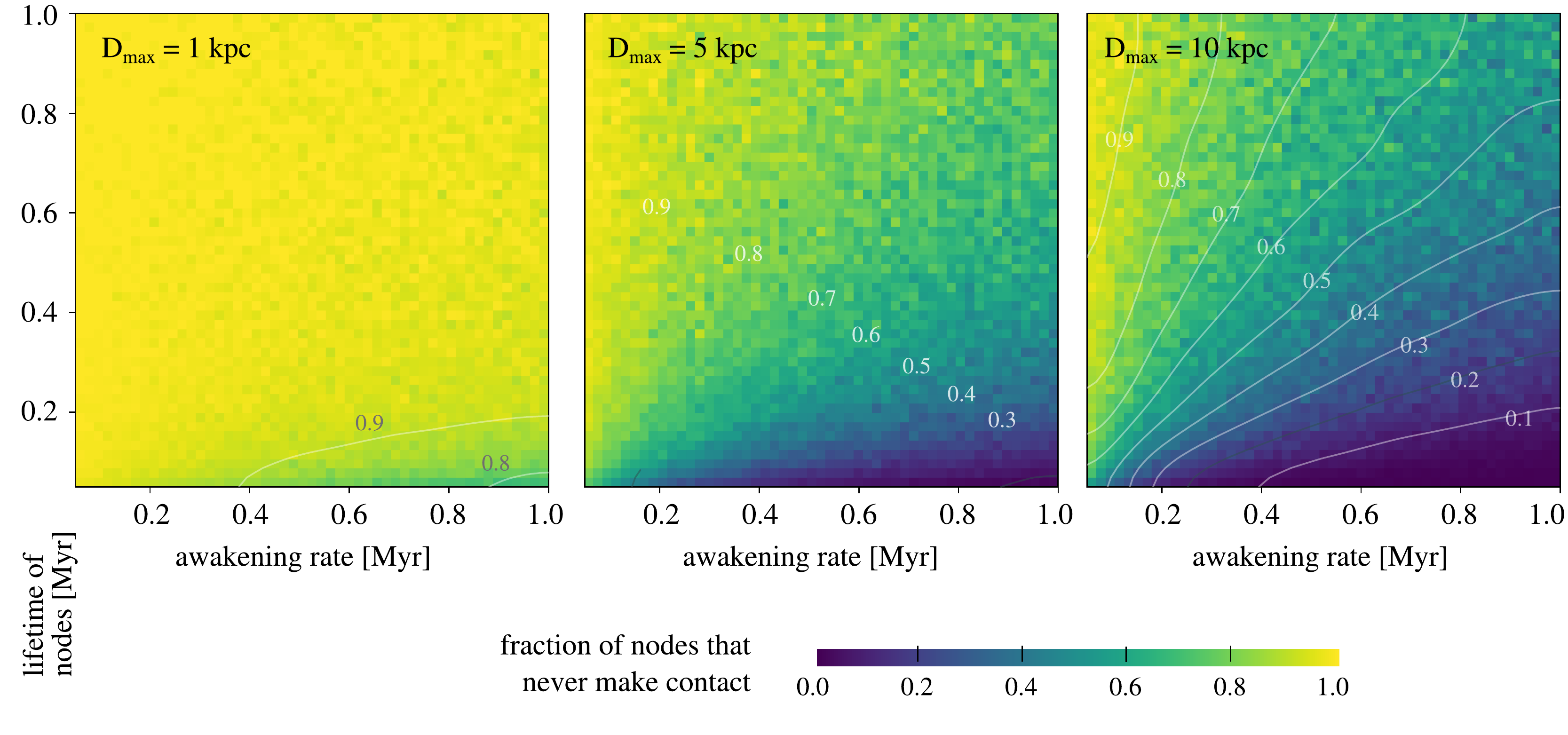}
   \caption{
The fraction of constrained causal contact nodes that never make contact (listening) as a
   function of $\tau_{\rm a}$ and $\tau_{\rm s}$, for 
$D_{\rm max}$=1~kpc (left panel),
$D_{\rm max}$=5~kpc (middle panel), and
$D_{\rm max}$=10~kpc (right panel).
The values of $\tau_{\rm a}$ and $\tau_{\rm s}$ are in the range
   5$\cdot10^5$ to 10$^6$~yr.
The matrix is shown as obtained from the simulations, and the level
   curves are shown for the smoothed matrix.
   }
   \label{F_never_contact}
\end{figure*}
 
\begin{figure*} % 2D color plot
   \centering
   \includegraphics[width=\textwidth]{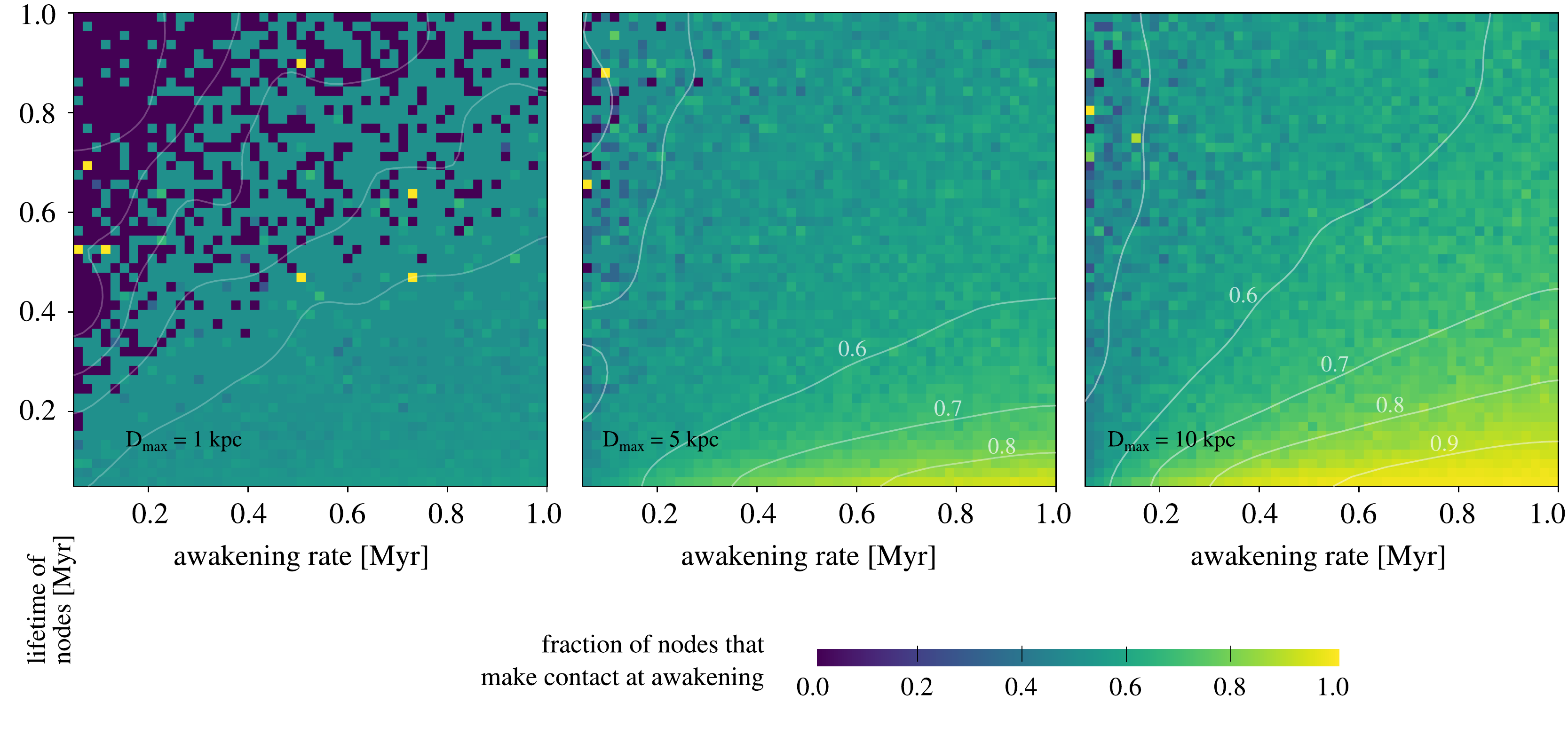}
   \caption{
The fraction of constrained causal contact nodes that make contact at the moment of the
   awakening (i.e., $t_{\rm A}=t_{\rm C}$), as a function of $\tau_{\rm a}$ and $\tau_{\rm s}$, for
$D_{\rm max}$=1~kpc (left panel),
$D_{\rm max}$=5~kpc (middle panel) and
$D_{\rm max}$=10~kpc (right panel).
The values of $\tau_{\rm a}$ and $\tau_{\rm s}$ are in the range
   5$\cdot10^5$ to 10$^6$~yr.
The matrix is shown as obtained from the simulations, and the level
   curves are shown for the smoothed matrix.
   }
   \label{F_C_at_A}
\end{figure*}

\subsection{Implementation of the SC3Net model}

% hearsay 
%
In order to make a reproducible project, we developed the tools that
allow to run the simulations and obtain the results shown in this
work.
The simulations were implemented on a Python-3 code, dubbed \hs{}
(Lares et al., in preparation), which is publicly accessible through
the GitHub platform\footnote{https://github.com/mlares/hearsay} under
the MIT-license.
The project is in the process of registration with the ‘’Astrophysics
Source Code Library’’ \citep[ASCL, ][]{2015JORS....3E..15A,
2020ASPC..522..731A}
From the viewpoint of a user, \hs{} is an object-oriented package that
exposes the main functionalities as classes and methods.
The code fulfills standar quality assurance metrics, that account for
testing, style, documentation and coverage.
In the configuration step, the user prepares the set of simulation
parameters through an initialisation file.
Since configuration files and simulation results are persistent, it is
straightforward to keep track of different experiments and the
experiments can be revisited easily.
An in-depth description of the methods can be found in the
documentation, which is automatically generated from \hs{} docstrings
and made public in the read-the-docs service
\footnote{https://hearsay.readthedocs.io/en/latest/}.
Since the simulation setup is configurable, the time required for a
simulation to complete depends on the simulation parameters.
It also depends on the hardware that is used to make the run, and on
whether the parallel option is set.
However, it is simple to make a local run of limited versions of the
experiments to have a sense of the time required by the code to
complete the simulations.
More information on this can be found on the documentation of the
software.

\begin{figure}
   \centering
   \includegraphics[width=0.5\textwidth]{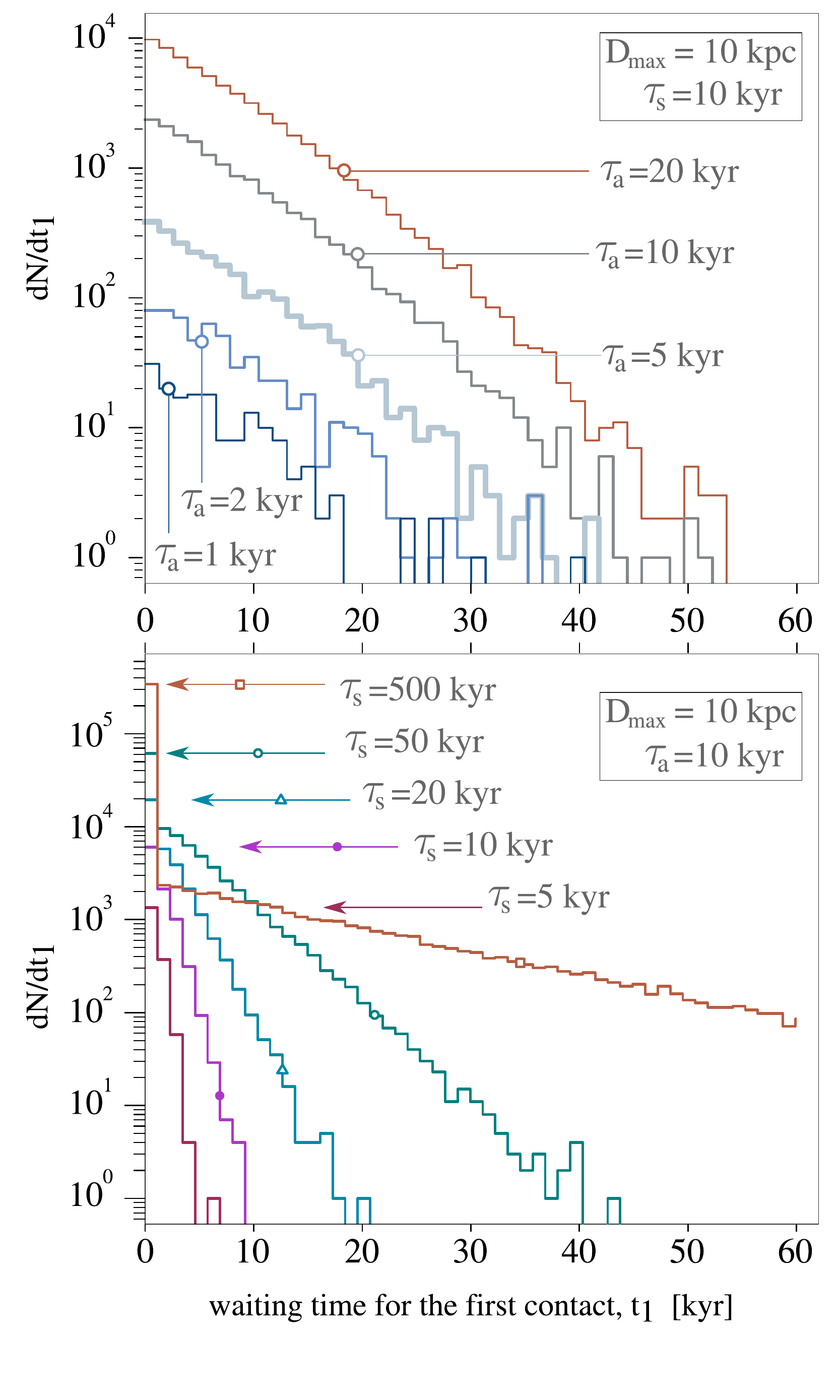}
   \caption{
Histograms of the mean waiting times for the first contact, for
several models.
   Upper panel shows the histograms for several values of $\tau_{\rm a}$, and
   $\tau_{\rm s}$=10~kyr and D$_{\rm max}$=10~kpc.
   Bottom panel shows the histograms for several values of $\tau_{\rm
   s}$, with $\tau_{\rm a}$=10~kyr and D$_{\rm max}$=10~kpc. The
   arrows indicate the sample and highlight the values of the first
   waiting time bin, which has a clear excess for the models with
   larger $\tau_{\rm s}$ values.
   The choice of these models was made in order to show the trends in
   the results as a function of the two temporal parameters.
   }
   \label{F_waiting_for_1C}
\end{figure}

\section{Results: exploring the parameter space}\label{S_results}

We implemented the simulation of a regular grids of models varying over
the parameter space, which covers 5204 models.
For each model, we simulated several realizations with different random
seeds, adding up a total of 
158546 simulation runs.
The number of random realzations varies from one (for the densest populated models)
up to 50 (for the sparsely populated models).
The parameters for the temporal aspects of the simulation (the mean
waiting time for the next awakening, $\tau_{\rm a}$, and the
mean lifetime, $\tau_{\rm s}$) cover the ranges
10$^2$-10$^6$~yr, with 
two linear partitions of 21 values (10$^2$-40500~yr) and 51 values
($5\cdot10^5$-10$^6$~yr)
for each parameter.
This partition was chosen after the requirement of the software to take linear bins, 
aiming at a better sampling of the low $\tau_{\rm a}$ and low
$\tau_{\rm s}$ region of the parameter space.
For the D$_{\rm max}$ parameter, we take the 
values 100~pc, 500~pc, 1~kpc, 5~kpc and 10~kpc.
\ttn{1}
In the Table \ref{T_simu_hypotheses} we show the three variable
parameters, the ranges of their values and the number of bins that
have been explored in the numerical experiments.
We also show the set of fixed parameters that take part in the simulation,
their values and the hypotheses that they represent.

As a product of the simulations, several quantities can be obtained.
Some of them are directly derived from the discrete events, namely,
the ID of emitting and receiving \cetis{} and the position in the
galaxy.
We also save the times of each of the events that are relevant to keep
track of the number of \cetis{} for each simulation, i.e., the times
of the four types of events.
The times of C-type and B-type events are used to derive the number of
contacts.
We also obtain quantities that represent the properties of the
\cetis{}, for example the total time elapsed between the A-type and
the D-type of each node, which represent their corresponding
lifetimes.
The time span of a \ceti{} listening another or being listened by
another node can also be derived by keepig track of the times of the
events in a simulation.
This way we can also compute the distribution in the simulated galaxy
of \cetis{} that reach contact, the distributions of the waiting times
until the first contact or the distributions of waiting times until
the next contact.
The following properties of the population of \cetis{} can also be
derived:
the fraction of the lifetime a node is listening to at least another
node (i.e., within their light cone),
the age of contacted \cetis{} at first contact,
the fraction of \cetis{} where the first contact is given at the
awakening,
the distribution of the number of contacts for each \ceti{},
the distribution of the number of contacts as a function of the age of
the node,
the number of contacts as a function of time in the galaxy,
the fractions of nodes that succeed in making contact,
and the distribution of distances between contacted nodes.
Another useful derived quantity is the duration of two-way
communication channels or the fraction of contacts that admit a
response.
It is also possible to analyze the relations between the distance to
\ceti{} vs. the time of two-way communications, the distance to
\ceti{} vs. the age of contacted node, the age of a node and the
maximum number of contacted nodes before the D--type event, or the
lifespan of a node vs. the maximum number of contacts.
All these quantities can be analyzed as a function of the simulation
parameters.

\subsection{Membership to the network of connected \cetis{}}\label{SS_members}

\ffn{3}
In Fig.~\ref{F_number_of_contacts} we show the empirical cumulative
distributions of the number contacts for \cetis{} in six different
samples, including short and long lifetimes, dense and sparse spatial
distribution and D$_{\rm max}$=10~kpc.
As it can be seen, the mean lifetime is more determinant than the mean
awakening rate (dense and sparse, represented by a different shade)
for the number of contacts.
The model with a dense awakening in the timeline (low $\tau_{\rm a}$)
maximizes the number of contacts, reaching a maximum of more than 300
contacts for a single \ceti{}.
This case, however, requires that a new node appears in the Galaxy
every 100 years on average.
Similarly, the model with long lifetimes has the maximum number of
contacts, reaching nearly 100 contacts for each \ceti{} in its entire
lifetime.
This is considerably larger than any model with a shorter lifetime,
which produce a number of contacts of at most the order of ten
contacts per \ceti{}.
It is expected then that a model where nodes appear with a high
frecuency and have very long lifetimes can reach tens of contacts on
the full time period between $t_{\rm A}$ and $t_{\rm D}$.
On the other side, a model where the activation of new nodes requires
a large waiting time and the survival time is short, contacts are
extremely rare.
We should point out that this plot has a logarithmic scale on the
x-axis, and it is the cumulative, not differential, empirical
distribution.
Therefore, the differences in the number of contacts for different
models are large.
This is a consequence of the wide range in both $\tau_{\rm a}$ and
$\tau_{\rm s}$ covered in the simulations.
With this ranges, the fraction of \cetis{} with no contact ranges from
nearly zero up to one.
This analysis is made in order to explore the behavior of the SC3Net
model.
\ffn{4}
On the Fig.~\ref{F_never_contact} we show 2D color maps with the
fraction of \cetis{} in the simulations that never make contact (i.e.,
never listen to another \ceti{}), as a function of the mean lifetime
($\tau_{\rm s}$, in the range 5$\cdot 10^4$ to 10$^6$ yr) and the mean
awakening time ($\tau_{\rm a}$, in the range 5$\cdot 10^4$ to 10$^6$
yr), for three different values of the maximum signal range, $D_{\rm
max}$=1~kpc (left panel), $D_{\rm max}=5$~kpc (middle panel), and
$D_{\rm max}=10$~kpc (right panel).
A clear pattern emerges, showing that the probability for a \ceti{} of
making causal contact with at least another \ceti{} during their
entire lifetime, increases with increasing $D_{\rm max}$, increasing
$\tau_{\rm s}$ and decreasing $\tau_{\rm a}$, following a roughly
linear dependence with the three parameters.
The results of the simulations that comprise the full range of values
for $\tau_{\rm a}$ and $\tau_{\rm s}$ from 5$\cdot10^2$~yr to $10^6$~yr
(not shown in the Figures) maintain a similar trend.
The number of \cetis{} that do not succeed in reaching the causal
contact regions of other nodes is a useful indicator of the degree of
isolation.
On the other hand, there is also the chance that a number of \cetis{}
are already in the causal contact region of other nodes, at the time
of their A-type events.
\ffn{5}
The fraction of \cetis{} that make the first contact at the awakening
event is shown in the Fig.~\ref{F_C_at_A}, as a function of the mean
lifetime ($\tau_{\rm s}$, in the range 5$\cdot 10^4$ to 10$^6$ yr) and
the mean awakening time ($\tau_{\rm a}$, in the range 5$\cdot 10^4$ to
10$^6$ yr).
The three panels correspond to different values of the maximum signal
range:
$D_{\rm max}=1$~kpc (left panel), $D_{\rm max}=5$~kpc (middle panel),
and $D_{\rm max}=10$~kpc (right panel).
The dependence of this metric with the three parameters is roughly
linear in all cases.

\subsection{Waiting time for a first contact}\label{SS_waiting}

In this subsection we analyze the distribution of the waiting times for
a first contact.
Such distribution can be considered to compute the probability for a
random \ceti{} to carry out the searching of other nodes and spend a given time until
the first contact is made, under the hypotheses of the experiment.
\ffn{6}
In the Fig.~\ref{F_waiting_for_1C} we show the histograms of the mean
waiting times for the first contact, for several models.
Upper panel (a) panel shows the histograms for several values of the
mean awakening time $\tau_{\rm a}$, with the mean survival time $\tau_{\rm
s}$ of 10~kyr.
Lower panel (b) shows the histograms for several values of $\tau_{\rm s}$,
with $\tau_{\rm a}$=10~kyr.
In both panels the value of D$_{\rm max}$ is set to 10~kpc.
As it can be seen, there is a clear trend where the number of nodes
that require a time $t_{\rm 1}$ to make the first contact 
decreases exponentially with the time $t_{\rm 1}$.
The fraction of the \cetis{} that have made at least one contact, as a
function of the elapsed time since the awakening, is computed with
respect to the total number of nodes that make causal contact at least
one time in the time range from the A--type up to the D--type events.
From a frequentist approach, the cumulative fraction of contacts from
the awakening ($t=$0~yr) up to a given time are related to the
estimation of the probability of listening during that time interval
and reaching to causal contact region of at least another node.
The complement of this value is the probability of observing in a time
interval with no success, i.e., without ever happening a C--type
event.
Clearly, this probability diminishes with time and tends to zero for
large time periods, meaning the a \ceti{} will eventually enter into
the light cone of other \cetis{}.
Remarkably, in densely populated models there is a clear excess at
$t_{\rm 1}$=0~yr.
That is, for a short period of time the initial moment is the most
promising for making a (causal) contact, for a given technology.
This is given by the fact that at the awakening event, many nodes are
already on the light cones of other nodes, so that the awakening time
offers the best chance of making contacts.
This offers a new approach to SETI programs, where the search for new
communication technologies or the exploration of new communication
channels has a fundamental role and could be more efficient than long
observation programs.

\section{Discussion}\label{S_discussion}

We have presented a stochastic model (SC3Net) to analyse the network
of constrained causally connected nodes in a simplified Galaxy.
It represents an idealised scenario of perfectly efficient emitters
and receivers with the restriction of a maximum distance separation.
These emitters and receivers correspond to nodes that form a
communication network whose properties depend on their density and
mean survival time.
The statistical analysis of the model allows to estimate the
probabilities for a random node and for a given model instance of
making contacts with other nodes, along with the waiting times and
durations of such contacts.
Using numerical simulations, we implemented this model to explore the
three-dimensional parameter space, considering a grid of mean time
separations between the activation of new nodes in the range
500--10$^6$~yr.
Additionally, we explore a grid of the mean survival times between
500~yr and 10$^6$~yr, and values for the maximum distance range of
signals of 100~pc, 500~pc, 1~kpc, 5~kpc and 10~kpc.
The simulation of each parameter point was performed several times
with different random seeds in order to improve the confidence of the
derived quantities.

Although the simulations use several hypotheses, we argue that the
model is not worthy of further complexity, given the limited knowledge
about the origin and persistence of life in the Galaxy.
The implementation of more detailed or sophisticated models would
increase the number of free parameters without any improvement in
their predictive power.
Thus, we take advantage of the simplicity of the model to explore the
parameter space in order to gain insight on the consequences of
different scenarios for the search of intelligent life.
Our analysis is not centred in obtaining the odds for the Earth to
make contact with another civilisation.
Instead, we focus on obtaining a statistical, parameter dependent
description of the possible properties of the communication networks
that comprise sets of nodes with broadcasting and reception
capabilities.
This causally connected nodes are sparsely distributed in both space
and time, making analytical treatments difficult and justifying the
simulation approach.

Under the hypotheses of our experiments, we conclude that a causal
contact is extremely unlikely unless the galaxy is densely populated
by intelligent civilisations with large average lifetimes.
This result is qualitatively similar to the results presented by
several authors, which state that a contact between the Earth and
another intelligent civilisation in the Galaxy is quite unlikely,
provided the maximum distance of the signal and the lifetime of the
emitter are not large enough.
This analysis supports the idea that, in order to increase the
possibilities of a contact, more active strategies of the emitter
would be required.
Some proposals in this direction include intertellar exploration,
colonization and settlement \citep{brin_great_1983, Dosovic2019,
galera_invasion_2019}, although it would require large temporal
scales.
\citet{Dosovic2019} use probabilistic cellular automaton simulations
to explore the parameter space of a model with colonization and
catastrophic events.
According to the timescales involved, their results could explain the
Fermi paradox.
Although our work does not take into account the colonization
hypothesis, it does consider catastrophic events implicitly in the
mean value and distribution of the lifetime, $\tau_{\rm s}$.
Other strategies could also increase the probability of contacts, for
example panspermia \citep[e.g.,][]{starling_virulence_2013} or
self--replicating probes \citep[e.g.,][]{barlow_galactic_2013},
although they would be too slow to make a significant impact on the
communication network among intelligent civilisations.
Our results are also consistent with those presented by
\citet{grimaldi_signal_2017}, who estimates an upper limit for the
mean number of extraterrestrial civilisations that could contact Earth
using Monte Carlo simulations, from a statistical model where the
width of the Galactic disk is not negligible.
Unlike most of the studies that make use of statistical models or
simulations \citep{cirkovic_temporal_2004, smith_broadcasting_2009,
bloetscher_using_2019}, our approach does not relay on the Drake
equation.
Thus, it does not need a detailed description or modeling of the physical
processes that give rise to intelligent life.
However, we argue that it is a valid empirical formulation to discuss
the probabilities of contact and the time scales involed in the
problem.
Ours is an alternative to the method proposed by
\citet{balbi_impact_2018}, who performs an analysis based on the Earth
with a different model.
Despite those differences, their results and ours are in general agreement.

We have also fond that there is a balance between the density of
\cetis{} and the mean lifetime since, as expected, a lower density can
be compensated by a longer active time period.
However, a large number of nodes does not easily compensate their
short lives to reach the same probability of causal contact than in
the case of a less populated galaxy but with very ancient
civilisations.
In all cases, for a short period of time (for instance, the time SETI
programs have been active on Earth), the maximum probability of making
a contact occurs at the moment of the awakening.
This suggests the possibility that an alternative SETI strategy could
be the search for alternative message carriers, for the case in which
the search has not been performed on the adequate channels.
Then, if a contact is produced for the first time, the origin of the
signal is more likely to be very old.
Also, the chances of entering the causal connected zone of a \ceti{}
does not grow linearly, but favours the first period of 10$^5$ years.

In the approach of this work, we have used computer simulations to address
the problem of the probabilities of causal contacts between locations
in the Galaxy with the possibility of sending and receiving messages.
Instead of making a number of assumptions, we have explored the
parameter space, reducing the problem to only three parameters and a
few simple hypotheses to perform a complete model for the population and
communication network in the galaxy.
This allows to consider the Fermi paradox from a new perspective, and
to propose an alternative treatment for the number of intelligent
emitter/receivers.
If the time intervals between the rise and fall of civilisations are short
compared to the time required for an electromagnetic signal to travel the 
large distances in the Galaxy, 
then the number of contacts would be limited to a low number.

The short time interval between the
rise and fall of civilisations, compared to the age and extension of
our Galaxy, is a fundamental limitation for the number of contacts.
The temporal dimension, which is missing in the Drake
equation, is a key factor to understand the network of contacts on
different scenarios.

\ack[Acknowledgement]{
This work was partially supported by the Consejo Nacional de
Investigaciones Cient\'{\i}ficas y T\'ecnicas (CONICET, Argentina),
the Secretar\'{\i}a de Ciencia y Tecnolog\'{\i}a, Universidad Nacional
de C\'ordoba, Argentina, and the Universidad Cat\'olica de C\'ordoba,
Argentina.
This research has made use of NASA's Astrophysics Data System.
Plots and simulations were made with software developed by the authors
the python language. 
Plots were postprocessed with inkscape.
Simulations were run on the Clemente cluster at the Instituto de
Astronom\'{\i}a Te\'orica y Experimental (IATE).
}

\ack[Disclosure statement]{
No competing financial interests exist.
}

\bibliographystyle{mn2e}
\setlength{\bibsep}{0.0pt}
%\bibliography{biblio_seti, biblio_software}

\end{document}